\documentclass[10pt,aps,prx,twocolumn,nofootinbib,superscriptaddress,floatfix]{revtex4-2}

\usepackage[utf8]{inputenc} 
\usepackage{graphicx} 
\graphicspath{{Figures/}} 
\usepackage{subcaption}
\usepackage{amsmath,amssymb} 

\usepackage{xcolor} 
\usepackage{booktabs} 
\usepackage{multirow} 
\usepackage{physics} 
\usepackage{siunitx} 
\usepackage{adjustbox} 

\usepackage{tikz}
\usetikzlibrary{arrows.meta, fit, backgrounds, decorations.pathreplacing, calc, positioning}
\usepackage{quantikz}
\usepackage{adjustbox}
\usepackage{braket}

\usepackage[colorlinks=true]{hyperref} 

\makeatletter
\let\label\ltx@label
\makeatother

\usepackage{cleveref} 
\crefname{figure}{figure}{figures}
\Crefname{figure}{Figure}{Figures}
\crefname{table}{table}{tables}
\Crefname{table}{Table}{Tables}
\crefname{section}{section}{sections}
\Crefname{section}{Section}{Sections}
\crefname{equation}{equation}{equations}
\Crefname{equation}{Equation}{Equations}

\begin{document}
\title{Hybrid Quantum-Classical Logistic Regression for Calibrated Classification of Pulsar Candidates}

\author{Chanelle Matadah Manfouo}
\email{Corresponding author: cmatadah@aimsric.org}
\affiliation{African Institute for Mathematical Sciences Research and Innovation Centre, Kigali, Rwanda }
\affiliation{Department of Physics, Stellenbosch University, Stellenbosch, 7600, South Africa}

\author{Donovan Slabbert}
\email{donovanslab@mweb.co.za}
\affiliation{Department of Physics, Stellenbosch University, Stellenbosch, 7600, South Africa}
\affiliation{National Institute for Theoretical and Computational Sciences (NITheCS),
Stellenbosch, 7600, South Africa}

\author{Prince Koree Osei}
\email{posei@nexteinstein.org}
\affiliation{African Institute for Mathematical Sciences Research and Innovation Centre, Kigali, Rwanda }
\affiliation{African Institute for Mathematical Sciences, Accra, Ghana}

\author{Francesco Petruccione}
\email{petruccione@sun.ac.za}
\affiliation{National Institute for Theoretical and Computational Sciences (NITheCS),
Stellenbosch, 7600, South Africa}
\affiliation{School of Data Science and Computational Thinking, Stellenbosch University,
Stellenbosch, 7600, South Africa}
\affiliation{Department of Physics, Stellenbosch University, Stellenbosch, 7600, South Africa}

\date{\today}

\begin{abstract}
\noindent

Reliable pulsar candidate ranking requires probability estimates that are not only discriminative but also well calibrated. We evaluate hybrid quantum-calssical logistic regression on the imbalanced \texttt{HTRU-$2$} dataset using three quantum feature encodings: angle encoding, amplitude encoding, and data re-uploading. The models are trained using analytic gradients and compared with classical baselines and a quantum support vector machine reference model under a paired-seed protocol. Evaluation combines rare-event discrimination, low-false-positive-rate recovery, probability calibration, and runtime analysis. Angle encoding gives the strongest performance among the quantum logistic regression variants. At shallow depth, the angle-encoded model remains close to the best classical baselines in discrimination and low-false-positive-rate recovery, while also giving the lowest calibration error at the benchmark configuration. Murphy decomposition shows that the angle-encoded model maintains low reliability error and high, stable resolution across circuit depths and training-set sizes. This means that its probability estimates preserve both calibration and meaningful separation between candidate groups. Data re-uploading is competitive at small depth but loses discrimination and resolution at larger depth in the present multi-qubit implementation, while amplitude encoding remains weaker across dataset sizes. 
Shallow angle-encoded quantum logistic regression therefore gives the best balance among the tested quantum logistic models, although simulation runtime remains a practical limitation.

\end{abstract}

\maketitle
\thispagestyle{plain}

\section{Introduction}\label{sec:intro}

Modern radio sky surveys produce extremely large observational datasets across multiple wavelength bands \cite{banks2021go, keane2014cosmic}.
The scale and complexity of the data impose substantial demands on analysis pipelines. For instance, next-generation facilities such as the Square Kilometre Array are expected to operate at exabyte-scale data rates \cite{kramer2015pulsar}.

Among the astronomical sources targeted by these surveys are pulsars. Pulsars are rapidly rotating, highly magnetised neutron stars that emit beams of electromagnetic radiation from their magnetic poles, producing periodic pulses when the beam sweeps across the Earth’s line of sight \cite{lorimer2005handbook}. Their exceptional rotational stability enables high-precision timing experiments that probe fundamental physics, including dense-matter physics and strong-field gravity \cite{foster1990constructing, hobbs2010international, verbiest2016international}. Modern radio surveys therefore devote substantial effort to the identification of new pulsars within large sets of candidate signals extracted from telescope observations.

Large surveys such as the High Time Resolution Universe (HTRU) survey generate vast amounts of candidate signals, the majority of which correspond to noise or radio-frequency interference \cite{lyon2016fifty}. Specifically, the \texttt{HTRU-$2$} \cite{lyon2016fifty} dataset, widely used as a benchmark for pulsar candidate classification, contains approximately 
$9\%$ pulsars. This skewed distribution complicates candidate screening because operational follow-up decisions rely on predicted probabilities rather than raw accuracy. Miscalibrated predictions can lead either to wasted telescope time on false positives or to the loss of genuine pulsars that are ranked too low for follow-up. Reliable calibration of predicted probabilities is therefore essential to ensure that prioritised candidates correspond to the true likelihood of a pulsar observation.

To address this screening challenge, surveys increasingly rely on machine learning classifiers. Classical approaches such as support vector machines, ensemble classifiers, and boosted models achieve strong discrimination performance on HTRU datasets \cite{lyon2016fifty, wang2019hybrid, tariq2022adaboost}. Deep neural networks further improve classification performance, although at increased computational cost \cite{xiao2020pulsar}. Most existing studies evaluate these models primarily using discrimination metrics such as accuracy, precision, recall, the area under the receiver operating characteristic curve (ROC–AUC), and the area under the precision–recall curve (PR–AUC). 

Recent studies have also explored quantum machine learning models for pulsar candidate classification. An early contribution by Kordzanganeh et al. proposed a single-qubit quantum neural network for pulsar candidate classification, where Quantum Asymptotically Universal Multi-Feature encoding maps scaled input features into repeated single-qubit rotations interleaved with trainable layers \cite{kordzanganeh2021quantum}. Subsequent work compared architectures such as quantum support vector machines and quantum convolutional neural networks, highlighting trade-offs between training efficiency and noise robustness \cite{slabbert2024pulsar}. Other studies investigated ensembles of shallow variational circuits to improve generalisation without increasing circuit depth \cite{mcfarthing2024classical}. Recent implementations using Qiskit further examined how encoding strategies and circuit design influence classification performance \cite{souza2025qiskit}.
Although these studies report promising results, their evaluation primarily focuses on discrimination metrics. Across both classical and quantum approaches, little attention has been given to the calibration of predicted probabilities, despite its direct relevance for ranking candidates in telescope follow-up pipelines.

In this work, Quantum Logistic Regression (QLR) is studied on the pulsar candidate classification task using the \texttt{HTRU-$2$} dataset. Logistic regression models the log-odds of class membership as a linear function of input features and remains a widely used probabilistic classifier in scientific applications, including pulsar candidate identification in radio astronomy \cite{lyon2016fifty}. In the formulation considered here, a parameterised quantum circuit generates a nonlinear feature representation of the input data through expectation values of measured observables. These quantum features are supplied to a classical logistic regression model that produces probability estimates for the pulsar class. Three encoding strategies are compared: angle encoding, amplitude encoding, and data re-uploading \cite{perez2020data}. 
The data re-uploading variant uses a direct multi-qubit circuit, where the eight input features are reintroduced across variational layers on an eight-qubit register. 
This differs from earlier \texttt{HTRU-$2$} quantum classifiers based on single-qubit reuploading or ensembles of single-qubit circuits \cite{kordzanganeh2021quantum,mcfarthing2024classical}.

For each encoding strategy, the resulting hybrid model jointly optimises the circuit parameters that define the quantum feature map and the weights of the logistic regression layer using a binary cross-entropy objective. This probabilistic formulation is important for pulsar searches because candidates must be ranked for follow-up, not only assigned binary labels. Reliable probability estimates are therefore central to candidate prioritisation, especially under high class imbalance. Consequently, this study examines QLR under imbalanced observational data. Model predictions are assessed using both discrimination metrics and calibration diagnostics. The aim is to determine whether shallow quantum feature models can achieve competitive discrimination for pulsar candidate classification while preserving reliable probability estimates for candidate ranking.

The rest of the paper is organized as follows. \Cref{sec:background} introduces binary classification and outlines classical logistic regression and its quantum counterpart. \Cref{sec:methods} describes the dataset, preprocessing, circuit architecture, and evaluation metrics while \cref{sec:results} presents the empirical analysis. \Cref{sec:conclusion} concludes the work.

\section{Background}\label{sec:background}

This section introduces the probabilistic framework underlying the classification task and presents the classical and quantum formulations of logistic regression considered in this work.

\subsection{Binary Classification and Logistic Regression}
\label{subsec:blr}

Binary classification is a supervised learning method that estimates the probability that an input sample belongs to one of two possible classes \cite{bishop2006pattern,hastie2009elements}. Formally, the task consists of learning a function $f:\mathbb{R}^{d} \rightarrow [0,1]$ such that $f(x) \approx p(y=1 \mid x)$, where $x \in \mathbb{R}^{d}$ denotes a $d$-dimensional input feature vector and $y \in \{0,1\}$ the corresponding class label. In the pulsar classification task considered here, this probability represents the likelihood that a candidate signal originates from a pulsar. The model is trained on a dataset $\{(x_i,y_i)\}_{i=1}^{N}$ and learns parameters that approximate this probability from observed examples.

A widely used parametric model for this task is logistic regression \cite{bishop2006pattern,hastie2009elements}. Logistic regression models the conditional probability of the positive class by assuming that the log-odds of the probability are a linear function of the input features. Specifically,
\begin{align}
\log\!\left(
\frac{p(y=1\mid x)}
     {1 - p(y=1\mid x)}
\right)
&= w^\top x + b
\label{eq:log_odds}
\\
p(y=1\mid x)
&= \sigma(w^\top x + b)
\label{eq:sigmoid}
\\
\sigma(z)
&= \frac{1}{1+\exp(-z)}
\label{eq:sigmoid_def}
\end{align}
where $w\in\mathbb{R}^{d}$ denotes the weight vector and $b\in\mathbb{R}$ the bias. The linear term $w^\top x + b$ defines the log-odds of the positive class, and the sigmoid function maps this quantity to a probability in the interval $[0,1]$.

Model parameters are estimated by minimising the binary cross-entropy loss, with $p_i = p(y=1\mid x_i)$,
\begin{equation}
\label{eq:binary-entropy}
\mathcal{L}(w,b)
=
-\sum_{i=1}^{N}
\left[
y_i \log p_i
+
(1-y_i)\log(1-p_i)
\right],
\end{equation}
which corresponds to maximising the likelihood of the observed labels under a Bernoulli model \cite{bishop2006pattern}.

\subsection{Quantum Logistic Regression}
\label{subsec:qlr}

QLR extends classical logistic regression by replacing the classical feature representation with a trainable feature map generated by a parameterised quantum circuit. Let $x \in \mathbb{R}^d$ denote an input feature vector. The circuit prepares an $n$-qubit quantum state according to
\begin{equation}
\ket{\Psi(x;\theta)}
=
V(\theta)\,U_{\mathrm{enc}}(x)\ket{0}^{\otimes n},
\end{equation}
where $U_{\mathrm{enc}}(x)$ is a data-dependent unitary that encodes the input features and $V(\theta)$ is a variational unitary with trainable parameters $\theta$ \cite{benedetti2019parameterized,cerezo2021variational}.

Expectation values of Hermitian observables measured on the state $\ket{\Psi(x;\theta)}$ define a set of quantum features
\begin{equation}
z_q(x;\theta)
=
\bra{\Psi(x;\theta)} O_q \ket{\Psi(x;\theta)},
\end{equation}
where $\{O_q\}_{q=1}^{m}$ denotes a collection of measurement operators. Collecting these measurements yields a feature vector $z(x;\theta) \in \mathbb{R}^m$. In the implementation used in this work, the observables correspond to single-qubit Pauli-$Z$ operators.

These quantum features are then supplied to a classical logistic regression model. The decision function is defined as
\begin{equation}
g(x;\theta,w,b)
=
w^\top z(x;\theta) + b,
\end{equation}
where $w \in \mathbb{R}^m$ and $b \in \mathbb{R}$ denote the weights and bias parameters. In this formulation, the quantum circuit acts as a nonlinear feature map, while the final classification stage corresponds to logistic regression applied to the learned quantum features.

The predicted probability of the positive class is obtained as
\begin{equation}
p(y=1 \mid x;\theta,w,b)
=
\frac{1}{1+\exp\!\left(-g(x;\theta,w,b)\right)}.
\end{equation}

The model parameters $(\theta,w,b)$ are estimated by minimising the binary cross-entropy loss defined in \cref{eq:binary-entropy}. The nonlinear mapping $x \mapsto z(x;\theta)$ arises from the data-dependent quantum circuit, while the logistic layer preserves the probabilistic interpretation of classical logistic regression.

Implementation details of the encoding strategy and circuit architecture used to construct $U_{\mathrm{enc}}(x)$ and $V(\theta)$ are provided in \cref{subsec:quantum_circuit}.

\section{Methods}\label{sec:methods}

This section describes the experimental procedure used to evaluate QLR on pulsar candidate classification.
The dataset and preprocessing are introduced, followed by the quantum circuit architecture, optimisation setup, paired-seed evaluation protocol, and performance metrics.

\subsection{Dataset and Preprocessing}\label{subsec:data}

Experiments are conducted on the \texttt{HTRU-$2$} dataset \cite{lyon2016fifty}, which contains pulsar candidate observations from the HTRU survey \cite{keith2010high}. Each sample is described by eight features derived from two diagnostic plots used in pulsar searches: the integrated pulse profile and the dispersion-measure signal-to-noise ratio (DM-SNR) curve. For each plot, four summary statistics are computed: mean, standard deviation, skewness, and excess kurtosis.
The dataset contains \(17{,}898\) samples, with \(1{,}639\) pulsars and \(16{,}259\) non-pulsars, corresponding to a class ratio of approximately \(1{:}10\), which corresponds to unbalanced data. Features are standardised to zero mean and unit variance using a StandardScaler fitted on the training set and applied to both training and test data. The scaling ensures consistent input ranges for data-dependent quantum operations and stabilises optimisation. A stratified split of \(80/20\) is used to preserve the class proportions in both subsets, as shown in \cref{tab:htru2_split}. All preprocessing steps are implemented using \texttt{scikit-learn} utilities \cite{pedregosa2011scikit}.

\begin{table}[!t]
\caption{Class distribution for the \(80\)/\(20\) stratified split of HTRU-\(2\).}
\centering
\begin{tabular}{lcccc}
\toprule
 & Total & \% of dataset & Train (80\%) & Test (20\%) \\
\midrule
Non–pulsars & 16{,}259 & 90.8\% & 13{,}007 & 3{,}252 \\
Pulsars     & 1{,}639  & 9.2\%  & 1{,}311  & 328 \\
\midrule
All         & 17{,}898 & 100\%  & 14{,}318 & 3{,}580 \\
\bottomrule
\end{tabular}
\label{tab:htru2_split}
\vspace{5pt}
\end{table}

\subsection{Quantum circuit model}\label{subsec:quantum_circuit}

\begin{figure*}
    \centering
    \includegraphics[width=\linewidth]{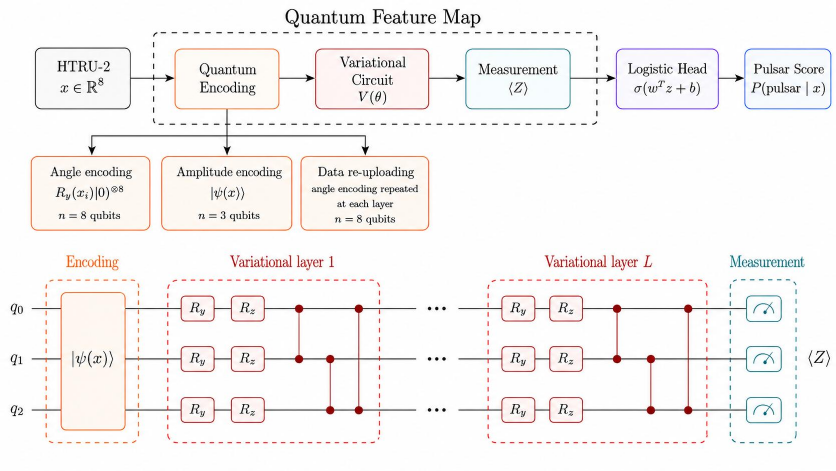}
    \caption{QLR pipeline for pulsar candidate classification and circuit architecture.
    The top panel shows the full classification pipeline. Classical features 
    $x \in \mathbb{R}^{8}$ from the \texttt{HTRU-2} dataset are encoded into a quantum state.
    The encoded state is processed by a variational circuit $V(\theta)$.
    Measurement produces a vector of Pauli-$Z$ expectation values 
    $z(x;\theta) \in [-1,1]^n$.
    These values are used as features in a logistic model, which outputs the probability 
    $P(\text{pulsar} \mid x)$ for candidate ranking.
    Three encoding strategies are considered.
    Angle encoding applies $R_y$ rotations proportional to each feature on $n=8$ qubits.
    Amplitude encoding maps the input vector to the amplitudes of a quantum state on $n=3$ qubits.
    Data re-uploading repeats angle encoding at each layer on $n=8$ qubits.
    The bottom panel shows the amplitude-encoding circuit architecture.
    The circuit contains an encoding block, repeated variational layers consisting of parametrised 
    $R_y$ and $R_z$ rotations with controlled-$Z$ entangling gates, and a measurement block 
    extracting Pauli-$Z$ expectation values.}
    \label{fig:qlr-pipeline}
\end{figure*}

The pipeline of the QLR framework considered is illustrated in \cref{fig:qlr-pipeline}. The \texttt{HTRU-$2$} dataset is encoded into a quantum state using, respectively, three encoding strategies. The resulting state is processed by a variational circuit $V(\theta)$ composed of parametrised layers $L$ followed by a measurement that produces a vector of Pauli expectation values $Z$$z(x;\theta) \in [-1,1]^n$. The late defines the learnt feature map. The feature vector is then passed to a logistic function to obtain the probability of the pulsar class. 

QLR operates on an input vector 
$x \in \mathbb{R}^8$ extracted from the \texttt{HTRU-$2$} dataset. 
A data-dependent unitary $U_{\mathrm{enc}}(x)$ prepares a quantum state from the reference state $\ket{0}^{\otimes n}$,
\begin{equation}
    \ket{\psi(x)} = U_{\mathrm{enc}}(x)\ket{0}^{\otimes n}.
\end{equation}

Three encoding strategies are considered within the same QLR framework: angle encoding, amplitude encoding, and data reuploading \cite{tacchino2019artificial,perez2020data,schuld2021machine}. 
These variants share the same variational circuit, measurement scheme, and training procedure, and differ only in how the classical input is embedded into the quantum state. 
For data reuploading, an alternating pairwise entanglement pattern is used, whereas angle and amplitude encoding employ a ring topology.

Angle encoding uses an eight-qubit register ($n=8$). Each feature $x_i$ parametrises a single-qubit rotation,
\begin{equation}
    U_{\mathrm{enc}}(x) =
    \bigotimes_{i=1}^{8} R_Y(\alpha x_i),
\end{equation}
where $\alpha$ is a scaling factor. The input is encoded once at the beginning of the circuit.

Amplitude encoding uses a three-qubit register ($n=3$). 
The input vector $x \in \mathbb{R}^8$ is normalised and embedded into the amplitudes of the quantum state,
\begin{equation}
    \ket{\psi(x)} =
    \sum_{i=0}^{7} \tilde{x}_i \ket{i},
\end{equation}
with $\sum_i |\tilde{x}_i|^2 = 1$.

QLR with data reuploading (QLR-DR) uses an eight-qubit register ($n=8$) and extends angle encoding by reintroducing the input at each layer of the circuit \cite{perez2020data}.  
For layers $l = 1,\dots,L-1$, the circuit applies data encoding followed by trainable rotations and an entangling block.  
The final layer applies data encoding followed by trainable rotations only, without an additional entangling operation.  

After state preparation, the system is evolved through a parametrised circuit of depth $L$, where each layer applies local rotations followed by entangling operations.

The parametrised unitary $V(\theta)$ consists of $L$ layers. 
For layer $l$, the trainable operation on qubit $q$ is
\begin{equation}
    U_q^{(l)} =
    R_Y(\theta^{(l)}_{q,1})\,R_Z(\theta^{(l)}_{q,2}),
\end{equation}
and the rotation block is
\begin{equation}
    U_{\mathrm{rot}}^{(l)} =
    \prod_{q=0}^{n-1} U_q^{(l)}.
\end{equation}

Entanglement is introduced using controlled-$Z$ gates.  
For angle and amplitude encoding, a ring topology is used,
\begin{equation}
    U_{\mathrm{ent}}^{(l)} =
    \prod_{q=0}^{n-1} \mathrm{CZ}(q,(q+1)\bmod n).
\end{equation}
For QLR-DR, an alternating pairwise pattern is used across layers.

The full transformation is written as
\begin{equation}
    \ket{\Psi(x;\theta)} =
    \left(
    \prod_{l=1}^{L}
    U_{\mathrm{ent}}^{(l)} U_{\mathrm{rot}}^{(l)}
    \right)
    U_{\mathrm{enc}}(x)\ket{0}^{\otimes n}.
\end{equation}

The expectation values of Pauli-$Z$ observables define the learned feature map. 
For each qubit $q$, the measurement yields
\begin{equation}
    z_q(x;\theta) =
    \langle \Psi(x;\theta) |
    Z_q
    | \Psi(x;\theta) \rangle,
\end{equation}
and collecting these values over all qubits defines the feature vector $z(x;\theta) \in [-1,1]^n$.

This construction yields a trainable quantum feature map $x \mapsto z(x;\theta)$, where the circuit architecture determines how input information is transformed before measurement.

\subsection{Optimisation and Training Protocol}\label{subsec:training}

The QLR model is trained by jointly optimising the quantum parameters $\theta$ and the classical parameters $(w,b)$.

For an input $x$, the circuit produces a feature vector 
$z(x;\theta) \in [-1,1]^8$, which is passed to a linear classifier,
\begin{equation}
g(x;\theta,w,b)
=
w^\top z(x;\theta) + b.
\end{equation}
Class probabilities are obtained through the logistic function defined in \cref{eq:sigmoid}, and training minimises the binary cross-entropy loss given in \cref{eq:binary-entropy}.

Optimisation is performed within a hybrid quantum-classical loop.  
Gradients with respect to $\theta$ are computed using the parameter-shift rule \cite{mitarai2018quantum, schuld2019evaluating}, which provides analytic derivatives of the expectation values for gates generated by two-eigenvalue operators.  
Gradients with respect to $(w,b)$ are obtained by automatic differentiation.

All experiments are implemented in PennyLane \cite{bergholm2018pennylane} using the \texttt{lightning.qubit} statevector simulator. The parameters are updated using Adam optimiser \cite{kingma2014adam} with the learning rate $0.01$ and the batch size $16$.  Early stopping is used to stabilise training, and termination is triggered if validation loss does not decrease for $15$ consecutive epochs. All optimisation hyperparameters are fixed across encoding strategies and circuit depths. The choice of hyperparameters was based on several tests with different batch sizes and learning rates, and the those who showed good stability and good training convergence of the loss were adopted.

\subsection{Paired-seed evaluation and \texorpdfstring{$\Delta$}{Delta}-metric computation}\label{subsec:paired_seed}

All experiments adopt a paired-seed evaluation protocol to enable statistically controlled comparisons between quantum and classical models. In this setup, each random seed defines one complete experimental replicate. Matched evaluation designs reduce variance arising from stochastic effects and are standard practice in comparative studies of learning algorithms \cite{dietterich1998approximate, demvsar2006statistical}.  Recent work further shows that uncontrolled random seeds can substantially affect reported performance \cite{schader2024don}.
For each seed $s$, a stratified \(80/20\) train/test split is generated, and all models are trained and evaluated on that split. 
The same seed also controls model initialisation and any stochastic optimisation procedures.  
This paired design reduces variability attributable to data sampling and optimisation randomness, allowing performance differences to be attributed primarily to model design.

For each seed and each evaluation metric $M$, the performance difference per seed is defined as
\begin{equation}
\Delta_s
=
M_{\mathrm{QLR},s}
-
M^{\ast}_{\mathrm{baseline},s},
\end{equation}
where $M_{\mathrm{QLR},s}$ denotes the metric value achieved by the quantum model and $M^{\ast}_{\mathrm{baseline},s}$ denotes the classical baseline that performs best for the same seed and metric.  
Selecting the best classical baseline on a per-seed, per-metric basis yields a conservative estimate of the performance of the quantum model.  
Positive values of $\Delta_s$ indicate that the quantum model outperforms the strongest classical baseline for the given seed and metric.

The $\Delta$ values are aggregated in the $N$ seeds as
\begin{equation}
\overline{\Delta}
=
\frac{1}{N}\sum_{s=1}^{N}\Delta_s,
\quad
\sigma_{\Delta}
=
\sqrt{\frac{1}{N-1}
\sum_{s=1}^{N}
(\Delta_s - \overline{\Delta})^2 }.
\end{equation}

To quantify the uncertainty of the mean performance difference, the standard error (SE) is defined as
\begin{equation}
\mathrm{SE}_{\Delta}
=
\frac{\sigma_{\Delta}}{\sqrt{N}},
\end{equation}
where $\sigma_{\Delta}$ denotes the standard deviation across the seeds and $N$ is the number of seeds. 
The standard error estimates the uncertainty associated with the mean value $\overline{\Delta}$ \cite{hastie2009elements}. 
The results are reported as $\overline{\Delta} \pm \mathrm{SE}_{\Delta}$, representing the mean paired performance gap and the uncertainty of this estimate across seeds.

\subsection{Classical Baselines and Comparative Quantum Model}
\label{subsec:baselines}

All baseline models are trained and evaluated under the paired-seed protocol described in \cref{subsec:paired_seed}.  For each seed, the classical baselines are trained on the same training subset used by the corresponding quantum model and evaluated on the same stratified test split.  Hyperparameters are selected separately for each seed using a grid search with cross-validation $3$ times in the training subset. The selection of models is based on average precision (AP), calculated using \texttt{average \_precision \_score}.  Average precision is used because the positive class is under-represented and the practical task is to rank likely pulsar candidates for follow-up. Therefore, it provides a selection criterion that is sensitive to the ordering of rare positive examples \cite{saito2015precision}.  The final comparison is not restricted to this selection criterion.  Models are evaluated using discrimination metrics, low-false-positive-rate operating points, calibration diagnostics, and runtime measurements.

Logistic regression is implemented using \texttt{scikit-learn} with $\ell_2$ regularisation and the \texttt{lbfgs} solver, with a maximum of $2000$ optimisation iterations. Class imbalance is handled using \texttt{class\_weight="balanced"}, which weights each class inversely to its frequency in the training subset. This weighting is applied only during model fitting and does not alter the class distribution of the test set. The inverse regularisation strength is tuned over $C \in \{0.1,\,1.0,\,10.0\}$.  For each seed, the selected configuration is refit on the full training subset before evaluation on the corresponding test split.

The second classical baseline is a Support Vector Machine with Radial Basis Function Kernel (SVM--RBF) with a radial basis function kernel \cite{cortes1995support}. Probability estimates are obtained by enabling \texttt{probability=True}, which applies Platt scaling to the SVM-RBF decision scores \cite{platt1999probabilistic}.  
Class imbalance is handled using balanced class weights.  
The regularisation parameter and kernel bandwidth are tuned jointly, with $C \in \{0.1,\,1.0,\,10.0\}$ and $\gamma \in \{\texttt{scale},\,\texttt{auto}\}$.  
For each seed, the selected configuration is refit on the full training subset and evaluated on the shared test split. Class imbalance is handled using \texttt{scale\_pos\_weight}, set to the ratio between negative and positive samples in the training subset. This weighting changes the training objective but leaves the test distribution unchanged.

The third classical baseline is extreme gradient boosting, implemented using XGBoost \cite{chen2016xgboost}.  
The binary logistic loss is used as the classification objective.  
Class imbalance is handled using \texttt{scale\_pos\_weight}, set to the ratio between negative and positive samples in the training subset.  
Hyperparameters are selected by grid search over the maximum tree depth $\{3,\,5\}$, learning rate $\{0.01,\,0.1\}$, and number of estimators $\{50,\,100\}$.  
Cross-validation again uses average precision as the selection criterion.  
The random state is fixed per seed, and \texttt{n\_jobs=1} is used to avoid additional variation from parallel execution.

In addition to the classical baselines, a quantum support vector machine (QSVM) is evaluated as a reference quantum model following the quantum kernel approach of Havlíček et al. \cite{havlivcek2019supervised}.  
Classical inputs are encoded into quantum states using an Instantaneous Quantum Polynomial-time feature map implemented through \texttt{IQPEmbedding}. The circuit is applied on $n$ qubits with a fixed number of repetitions. For two inputs $x$ and $x'$, the kernel is defined as the squared overlap between the corresponding quantum states,
\begin{equation}
k(x,x') = \left| \braket{\phi(x')}{\phi(x)} \right|^2 .
\end{equation}
The kernel matrices are evaluated using the \texttt{lightning.qubit} statevector simulator.  
The resulting kernel matrices are first symmetrised. If numerical eigenvalue errors make a kernel matrix non-positive-semidefinite, the matrix is projected onto the positive semidefinite cone before being passed to the classical SVM solver \cite{higham1988computing}. The corrected matrix is used as a precomputed kernel in a classical SVM with regularisation parameter $C=1.0$ and balanced class weights.

\begin{table}[t]
\centering
\caption{Hyperparameter search spaces used for the classical baseline methods and the details for the comparative QSVM mode.}
\label{tab:baseline_hyperparameters}
\begin{tabular}{lll}
\toprule
Model & Hyperparameter & Values \\
\midrule
Logistic regression 
& $C$ 
& $\{0.1,\,1.0,\,10.0\}$ \\

\addlinespace[0.35em]

SVM--RBF 
& $C$ 
& $\{0.1,\,1.0,\,10.0\}$ \\
& $\gamma$ 
& $\{\texttt{scale},\,\texttt{auto}\}$ \\

\addlinespace[0.35em]

XGBoost 
& Maximum depth 
& $\{3,\,5\}$ \\
& Learning rate 
& $\{0.01,\,0.1\}$ \\
& Number of estimators 
& $\{50,\,100\}$ \\

\addlinespace[0.35em]

QSVM 
& $C$ 
& $1.0$ \\
& Feature map 
& \texttt{IQPEmbedding} \\
& Number of repetitions 
& fixed \\
\bottomrule
\end{tabular}
\end{table}

\Cref{tab:baseline_hyperparameters} summarises the hyperparameter search spaces used to tune the classical baselines before comparison, together with the fixed settings used for the comparative QSVM model. All imbalance-handling choices are applied only during training. The held-out test split is neither reweighted nor resampled, so the reported metrics reflect the original class distribution. Training and prediction times are reported separately to distinguish the cost of kernel construction from the cost of inference.

\subsection{Performance Metrics}\label{subsec:metrics}

All models are evaluated using discrimination, operating-point, calibration, and runtime metrics.  
Precision-recall based metrics are included because the positive class is under-represented and the task requires reliable ranking of rare pulsar candidates \cite{davis2006relationship,saito2015precision}.  
Calibration is evaluated because the proposed model outputs probabilities intended to support candidate prioritisation rather than only binary decisions \cite{guo2017calibration,murphy1973new}.

At a fixed decision threshold, predictions are summarised by the confusion matrix entries: true positives (TP), false positives (FP), true negatives (TN), and false negatives (FN).  
Precision is defined as
\begin{equation}
\mathrm{Precision}
=
\frac{\mathrm{TP}}{\mathrm{TP}+\mathrm{FP}},
\end{equation}
and quantifies the fraction of predicted pulsar candidates that are true pulsars.  
Recall is defined as
\begin{equation}
\mathrm{Recall}
=
\frac{\mathrm{TP}}{\mathrm{TP}+\mathrm{FN}},
\end{equation}
and measures the fraction of true pulsars that are correctly identified.  
The F1 score is the harmonic mean of precision and recall.  
The false negative rate is defined as
\begin{equation}
\mathrm{FNR}
=
\frac{\mathrm{FN}}{\mathrm{TP}+\mathrm{FN}},
\end{equation}
and quantifies the fraction of true pulsars missed by the classifier.  
The false positive rate is defined as
\begin{equation}
\mathrm{FPR}
=
\frac{\mathrm{FP}}{\mathrm{FP}+\mathrm{TN}}.
\end{equation}

Threshold-free discrimination is evaluated using the area under the receiver operating characteristic curve (ROC--AUC) and PR--AUC.  
ROC--AUC summarises the trade-off between true positive rate and false positive rate across decision thresholds.  
In this work, PR--AUC denotes the average precision score computed using \texttt{average\_precision\_score}.  
PR--AUC is used as the principal ranking metric because it reflects the ordering of rare positive examples under class imbalance \cite{davis2006relationship,saito2015precision}.

To reflect operational constraints in large-scale pulsar surveys, performance is also evaluated at fixed false-positive-rate budgets.  
For each $\alpha \in \{0.01,0.05\}$, the threshold $\tau_\alpha$ is selected to maximise recall subject to
\begin{equation}
\mathrm{FPR}(\tau_\alpha)\le \alpha.
\end{equation}
The corresponding recall is reported as
\begin{equation}
\mathrm{Recall@FPR}\le\alpha
=
\mathrm{Recall}(\tau_\alpha).
\end{equation}
At $\alpha=0.01$, the associated precision and false negative rate are also reported.  
These operating-point metrics quantify pulsar recovery under strict false-positive budgets, where small false-positive rates can still correspond to large numbers of spurious candidates in large surveys \cite{lyon2016fifty}.

Calibration describes the agreement between predicted probabilities and empirical outcome frequencies. The Brier score is used as a proper scoring rule for probabilistic predictions \cite{glenn1950verification},
\begin{equation}
\mathrm{Brier}
=
\frac{1}{N}\sum_{i=1}^{N} (p_i-y_i)^2,
\end{equation}
where $p_i$ is the predicted probability of the positive class and $y_i\in\{0,1\}$ is the observed label. Lower Brier scores correspond to better probabilistic accuracy.

Murphy's decomposition is used to separate the Brier score into reliability, resolution, and uncertainty terms \cite{murphy1973new,murphy1977reliability,siegert2017simplifying},
\begin{equation}
\mathrm{Brier}
=
\mathrm{Reliability}
-
\mathrm{Resolution}
+
\mathrm{Uncertainty}.
\end{equation}
The reliability term measures calibration error.  
The resolution term measures how strongly the predicted probabilities separate cases with different empirical event frequencies.  
The uncertainty term depends only on the base rate
\begin{equation}
\bar{y}
=
\frac{1}{N}\sum_{i=1}^{N} y_i
\end{equation}
and reflects the baseline variability of the binary labels before using any model predictions.  
For binary labels, the uncertainty term is
\begin{equation}
\mathrm{Uncertainty}
=
\bar{y}(1-\bar{y}).
\end{equation}
The resolution term is bounded above by this uncertainty value, which provides a natural scale for interpreting resolution.

The expected calibration error (ECE) is also reported as a scalar summary of miscalibration \cite{guo2017calibration},
\begin{equation}
\mathrm{ECE}
=
\sum_{m=1}^{M}\frac{|B_m|}{N}
\left|
\mathrm{freq}(B_m)
-
\mathrm{conf}(B_m)
\right|,
\end{equation}
where predictions are partitioned into $M=15$ probability bins $B_m$.  
Here, $\mathrm{freq}(B_m)$ is the empirical fraction of positive labels in bin $B_m$, and $\mathrm{conf}(B_m)$ is the mean predicted positive-class probability in the same bin.  
The choice $M=15$ balances the resolution of the reliability curve with statistical stability for the dataset sizes considered here.

Reliability diagrams provide a visual diagnostic of calibration \cite{murphy1977reliability,niculescu2005predicting}. They plot the empirical positive-class frequency against the mean predicted probability across bins.  Perfect calibration corresponds to the diagonal.
Depth-dependent reliability and resolution curves are computed using the same probability bins as the reliability diagrams.  These curves characterise how circuit depth and training-set size affect probability calibration and class separation.

Runtime is measured as wall-clock time using high-resolution timers.  
Training time includes model fitting and, for QSVM, construction of the full training kernel matrix.  Prediction time includes forward inference and, for QSVM, computation of the test--train kernel matrix.  All experiments were executed on the same remote server.  GPU acceleration was not used, and all models were run in single-threaded CPU mode to improve comparability of runtime measurements.

\section{Results and Discussion}\label{sec:results}

QLR is evaluated on the \texttt{HTRU-$2$} pulsar dataset under the paired-seed protocol described in \cref{subsec:paired_seed}. For each seed, all models use the same stratified data split, which enables matched comparison across models. Three QLR variants are considered: angle encoding (QLR-angle), amplitude encoding (QLR-amplitude), and data re-uploading (QLR-DR). These variants are compared with Logistic Regression, SVM-RBF, and XGBoost as classical baselines. QSVM is included as a reference quantum model but is excluded from the $\Delta$-metric analysis. QLR-angle and QLR-DR use an eight-qubit register, while QLR-amplitude represents the eight input features using a three-qubit state.

The results address whether shallow QLR circuits can achieve competitive discrimination and reliable probability estimates under class imbalance with limited training data ($N \leq 1000$). Unless stated otherwise, values are reported as mean $\pm$ standard error over three independent seeds. The analysis is organised into three parts. First, a representative configuration, $N=1000$ and $L=3$, is used to compare predictive performance, calibration, and computational cost across all models. Second, the effect of circuit depth and training-set size is examined to assess how the QLR variants scale with $L$ and $N$. 
Third, probability calibration is analysed in more detail using ECE, reliability diagrams, and Murphy decomposition. This structure connects discrimination performance with calibration behaviour, since pulsar candidate selection requires probability estimates that can support ranking and telescope follow-up decisions. All numerical experiments were conducted remotely on a workstation running Ubuntu with the following hardware specifications: an Intel Core i$9$-$14900K$F processor with $24$ cores and $32$ threads, $125$ GB RAM, and an NVIDIA GeForce RTX $4090$ GPU. The GPU was not used in the present experiments, and all runs were executed using a single core.

\subsection{Benchmark Performance at $N=1000$, $L=3$}\label{subsec:main-model-results}

This subsection compares all models at a fixed training-set size of $N=1000$ and a circuit depth of $L=3$. This configuration provides a common reference point for evaluating discrimination performance, low-false-positive-rate screening, fixed-threshold error behaviour, probability calibration, and computational cost. 
The depth $L=3$ is used as a representative shallow-depth benchmark because QLR-angle is already stable in this range, while QLR-amplitude performs competitively among the tested depths. 
QLR-DR is also included at this depth so that all QLR variants are compared under the same benchmark setting. The full depth-dependent behaviour is analysed in \cref{subsec:scaling}.

\begin{table*}[!t]
\centering
\renewcommand{\arraystretch}{1.3}
\caption{
Predictive performance and operational screening metrics on \texttt{HTRU-$2$} at $N=1000$ and $L=3$. Values are reported as mean $\pm$ standard error over three independent seeds. Bold entries indicate the best value in each row.
}
\label{tab:performance_regime}
\begin{tabular}{lccccccc}
\toprule
Metric & LogReg & SVM-RBF & XGBoost & QSVM & QLR-angle & QLR-amplitude & QLR-DR \\
\midrule
PR-AUC        & $\mathbf{0.930 \pm 0.001}$ & $0.866 \pm 0.002$ & $0.910 \pm 0.005$ & $0.857 \pm 0.009$ & $0.910 \pm 0.002$ & $0.775 \pm 0.005$ & $0.907 \pm 0.005$ \\
ROC-AUC       & $\mathbf{0.972 \pm 0.002}$ & $0.959 \pm 0.001$ & $0.967 \pm 0.003$ & $0.959 \pm 0.003$ & $0.968 \pm 0.002$ & $0.889 \pm 0.001$ & $0.963 \pm 0.005$ \\
Recall@FPR1\% & $\mathbf{0.868 \pm 0.003}$ & $0.846 \pm 0.013$ & $0.857 \pm 0.008$ & $0.804 \pm 0.024$ & $0.866 \pm 0.006$ & $0.667 \pm 0.013$ & $0.846 \pm 0.003$ \\
Recall@FPR5\% & $\mathbf{0.926 \pm 0.001}$ & $0.904 \pm 0.009$ & $0.900 \pm 0.005$ & $0.909 \pm 0.003$ & $0.914 \pm 0.005$ & $0.727 \pm 0.008$ & $0.912 \pm 0.005$ \\
\bottomrule
\end{tabular}
\end{table*}

\begin{table*}[!t]
\centering
\renewcommand{\arraystretch}{1.3}
\caption{
Model calibration, fixed-threshold error characteristics, and computational cost on \texttt{HTRU-$2$} at $N=1000$ and $L=3$. Training and prediction times are reported in seconds. Values are reported as mean $\pm$ standard error over three independent seeds. Downward arrows indicate metrics for which lower values are better.
}
\label{tab:calibration_runtime}
\begin{tabular}{lccccccc}
\toprule
Metric & LogReg & SVM-RBF & XGBoost & QSVM & QLR-angle & QLR-amplitude & QLR-DR \\
\midrule

\multicolumn{8}{l}{\textit{Calibration}} \\
Brier $\downarrow$ & $0.037 \pm 0.005$ & $0.022 \pm 0.002$ & $0.027 \pm 0.002$ & $0.023 \pm 0.002$ & $\mathbf{0.018 \pm 0.001}$ & $0.037 \pm 0.001$ & $0.019 \pm 0.001$ \\
ECE $\downarrow$   & $0.097 \pm 0.010$ & $0.016 \pm 0.006$ & $0.039 \pm 0.010$ & $0.012 \pm 0.002$ & $\mathbf{0.008 \pm 0.002}$ & $0.033 \pm 0.003$ & $0.009 \pm 0.000$ \\

\midrule
\multicolumn{8}{l}{\textit{Error characteristics (threshold = 0.5)}} \\
Precision     & $0.766 \pm 0.042$ & $0.879 \pm 0.020$ & $0.816 \pm 0.027$ & $0.875 \pm 0.017$ & $0.942 \pm 0.005$ & $\mathbf{0.973 \pm 0.011}$ & $0.918 \pm 0.003$ \\
Recall        & $\mathbf{0.914 \pm 0.008}$ & $0.851 \pm 0.008$ & $0.876 \pm 0.009$ & $0.820 \pm 0.017$ & $0.820 \pm 0.022$ & $0.595 \pm 0.033$ & $0.826 \pm 0.003$ \\
FNR $\downarrow$ & $\mathbf{0.086 \pm 0.008}$ & $0.149 \pm 0.008$ & $0.124 \pm 0.009$ & $0.180 \pm 0.017$ & $0.180 \pm 0.022$ & $0.405 \pm 0.033$ & $0.174 \pm 0.003$ \\

\midrule
\multicolumn{8}{l}{\textit{Runtime (s)}} \\
Training   & Negligible & Negligible & Negligible & $709.55 \pm 1.04$ & $7505.11 \pm 90.35$ & $1035.55 \pm 1.30$ & $13170.15 \pm 2491.16$ \\
Prediction & Negligible & Negligible & Negligible & $5071.46 \pm 8.38$ & $9.38 \pm 0.17$ & $4.25 \pm 0.02$ & $11.85 \pm 3.00$ \\

\bottomrule
\end{tabular}
\end{table*}

\Cref{tab:performance_regime} reports discrimination and low-false-positive-rate screening metrics. Logistic Regression gives the highest PR-AUC and ROC-AUC, and also gives the best recall under both false-positive-rate constraints. QLR-angle remains close to the best classical model, with PR-AUC $0.910 \pm 0.002$, ROC-AUC $0.968 \pm 0.002$, and Recall@FPR1\% $0.866 \pm 0.006$. The corresponding Logistic Regression values are PR-AUC $0.930 \pm 0.001$, ROC-AUC $0.972 \pm 0.002$, and Recall@FPR1\% $0.868 \pm 0.003$. QLR-DR shows similar but slightly lower performance at this depth, especially under the stricter false-positive-rate constraint. QLR-amplitude is weaker across the discrimination and screening metrics, with Recall@FPR1\% reduced to $0.667 \pm 0.013$.

\Cref{tab:calibration_runtime} complements these discrimination results by reporting calibration metrics, fixed-threshold error characteristics, and computational cost. QLR-angle gives the lowest Brier score and ECE, indicating the lowest calibration error among the models compared at this configuration. QLR-DR is close to QLR-angle in both calibration metrics, with Brier score $0.019 \pm 0.001$ and ECE $0.009 \pm 0.000$. 
By contrast, Logistic Regression has strong discrimination but higher calibration error, with ECE $0.097 \pm 0.010$. This difference motivates reporting discrimination and calibration together for pulsar candidate ranking. A model can rank candidates well while assigning probability values that are less reliable as confidence estimates.

The fixed-threshold metrics in \cref{tab:calibration_runtime} describe a different operating regime from the low-FPR metrics in \cref{tab:performance_regime}. 
At the fixed threshold $0.5$, Logistic Regression gives the highest recall and lowest false-negative rate. QLR-angle has a lower recall and a higher false-negative rate at this threshold, despite strong Recall@FPR1\%. This difference occurs because the threshold $0.5$ is not necessarily the operating point that maximises pulsar recovery under a false-positive-rate constraint. The low-FPR metrics describe recovery under constrained false positives, while the fixed-threshold false-negative rate describes missed pulsars under the default decision threshold. Both views are relevant for pulsar screening, where follow-up resources are limited and missed pulsars reduce discovery completeness.

The runtime rows in \cref{tab:calibration_runtime} show that the favourable calibration of QLR-angle comes with a substantial training cost. The classical baselines are negligible on the time scale of the quantum simulations and are therefore reported as negligible in \cref{tab:calibration_runtime}. Among the quantum models, QSVM has the largest prediction time because prediction requires kernel evaluations against the training data. The QLR variants have much lower prediction times, but their training costs differ strongly by encoding. QLR-amplitude is the least expensive QLR variant to train, while QLR-DR is the most expensive. At $L=3$, QLR-DR requires $13170.15 \pm 2491.16$ seconds for training, compared with $1035.55 \pm 1.30$ seconds for QLR-amplitude and $7505.11 \pm 90.35$ seconds for QLR-angle. These results identify encoding choice as a central factor in the trade-off between predictive performance, calibration, and computational cost. The following sections examine whether these trade-offs persist as circuit depth and training-set size are varied.

\subsection{Scaling with Circuit Depth and Training-Set Size}
\label{subsec:scaling}

The benchmark analysis in \cref{subsec:main-model-results} fixes $N=1000$ and $L=3$. This subsection examines whether the same trends persist when the training-set size and circuit depth are varied. 
The scaling analysis focuses on PR-AUC, Recall@FPR1\%, ECE, and training time. These quantities connect discrimination performance, low-false-positive-rate screening, probability calibration, and computational cost.

The effect of training-set size is first studied at fixed circuit depth $L=3$, with $N \in \{200,500,1000\}$. \Cref{tab:qlr_data_scaling} reports PR-AUC and Recall@FPR1\% for the three QLR variants. This table isolates the effect of increasing the number of training samples while keeping the circuit architecture fixed.

\begin{table}[!t]
\centering
\caption{
QLR performance at fixed circuit depth $L=3$ across training-set sizes. 
Values are reported as mean $\pm$ standard error over three independent seeds.
}
\label{tab:qlr_data_scaling}
\begin{tabular}{llcc}
\toprule
Encoding & $N$ & PR-AUC & Recall@FPR1\% \\
\midrule
\multirow{3}{*}{QLR-angle}
  & 200  & $0.904 \pm 0.007$ & $0.845 \pm 0.007$ \\
  & 500  & $0.908 \pm 0.004$ & $0.850 \pm 0.015$ \\
  & 1000 & $0.910 \pm 0.002$ & $0.866 \pm 0.006$ \\
\midrule
\multirow{3}{*}{QLR-amplitude}
  & 200  & $0.723 \pm 0.012$ & $0.616 \pm 0.034$ \\
  & 500  & $0.752 \pm 0.011$ & $0.655 \pm 0.011$ \\
  & 1000 & $0.775 \pm 0.005$ & $0.667 \pm 0.013$ \\
\midrule
\multirow{3}{*}{QLR-DR}
  & 200  & $0.879 \pm 0.005$ & $0.790 \pm 0.010$ \\
  & 500  & $0.891 \pm 0.003$ & $0.802 \pm 0.008$ \\
  & 1000 & $0.907 \pm 0.005$ & $0.846 \pm 0.003$ \\
\bottomrule
\end{tabular}
\end{table}

QLR-angle shows limited variation as $N$ increases. 
PR-AUC changes from $0.904 \pm 0.007$ at $N=200$ to $0.910 \pm 0.002$ at $N=1000$, while Recall@FPR1\% changes from $0.845 \pm 0.007$ to $0.866 \pm 0.006$. At $L=3$, QLR-angle therefore gains only modestly from increasing $N$ over the tested range. QLR-DR follows a stronger data-scaling trend, with PR-AUC increasing from $0.879 \pm 0.005$ to $0.907 \pm 0.005$ and Recall@FPR1\% increasing from $0.790 \pm 0.010$ to $0.846 \pm 0.003$. QLR-amplitude also improves with additional data, but remains below the other QLR variants at all training-set sizes. At $N=1000$, its PR-AUC is $0.775 \pm 0.005$ and its Recall@FPR1\% is $0.667 \pm 0.013$. Across the tested training sizes, the ordering of the QLR variants remains unchanged: QLR-angle gives the strongest discrimination, QLR-DR remains close but lower, and QLR-amplitude gives the weakest discrimination.

\begin{figure*}[!t]
\centering
\begin{subfigure}[t]{\textwidth}
    \centering
    \includegraphics[width=\linewidth]{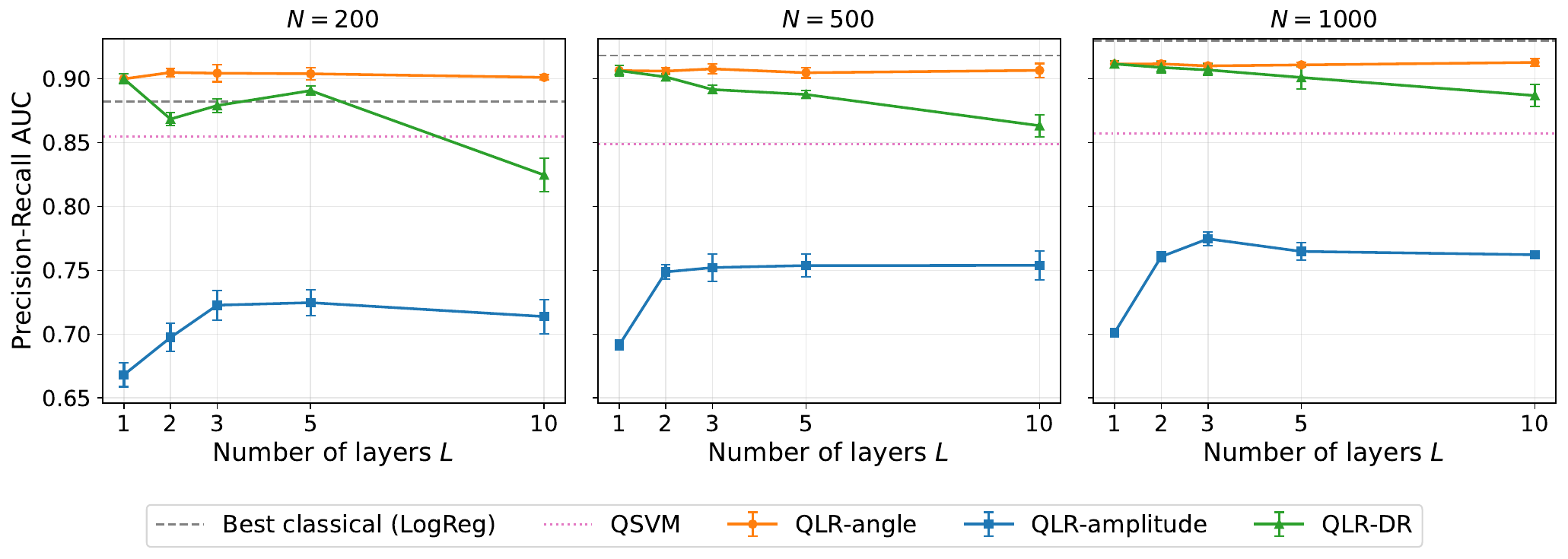}
    \caption{PR-AUC}
    \label{fig:depth_pr}
\end{subfigure}
\vspace{0.5em}
\begin{subfigure}[t]{\textwidth}
    \centering
    \includegraphics[width=\linewidth]{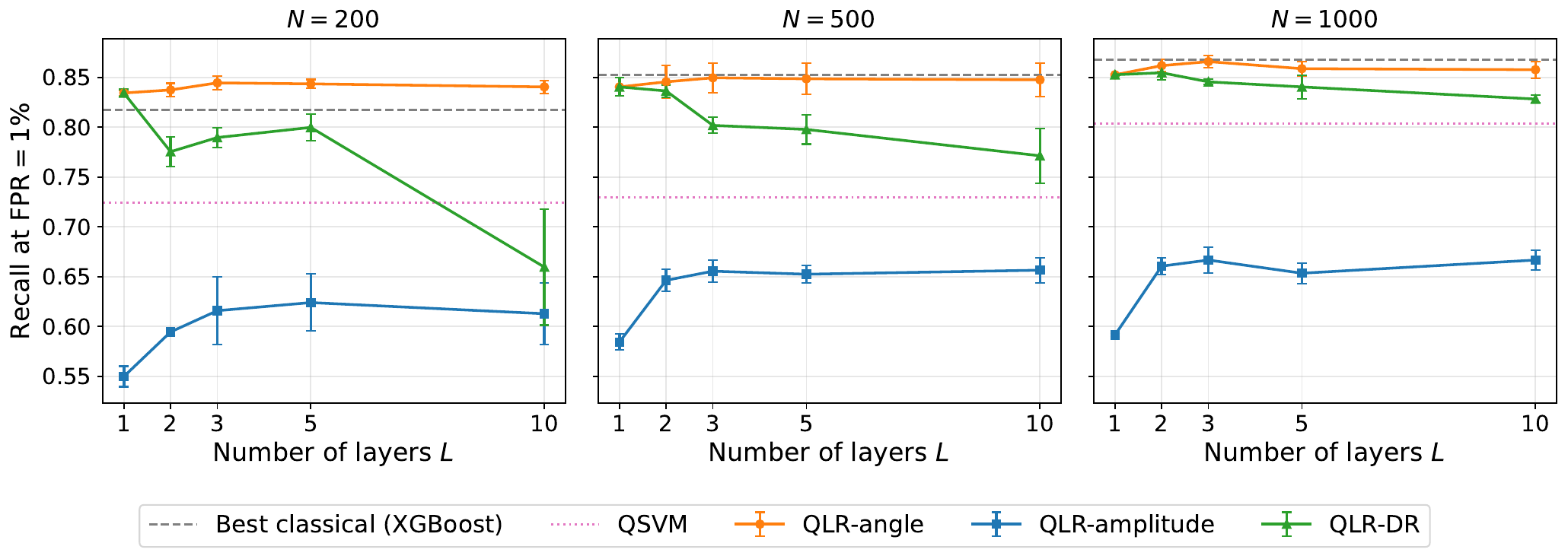}
    \caption{Recall@FPR1\%}
    \label{fig:depth_recall}
\end{subfigure}
\caption{
Depth dependence of QLR discrimination performance for $N \in \{200,500,1000\}$. 
The upper panel reports PR-AUC, and the lower panel reports Recall@FPR1\%. Each column corresponds to a specific training-set size. Error bars denote standard error over three independent seeds. Dashed horizontal lines indicate the best classical baseline for the corresponding metric and training size. Dotted horizontal lines indicate the QSVM reference value.
}
\label{fig:depth_scaling}
\end{figure*}

\Cref{fig:depth_scaling} examines the effect of circuit depth. 
The figure reports PR-AUC and Recall@FPR1\% as functions of $L \in \{1,2,3,5,10\}$ for $N \in \{200,500,1000\}$. 
The dashed horizontal line marks the best classical baseline for the corresponding metric and training size, while the dotted horizontal line marks the QSVM reference value. 
Error bars denote standard error over three independent seeds.

QLR-angle achieves high performance at $L=1$ and changes little as the depth increases. At $N=1000$, PR-AUC remains in the range from $0.909$ to $0.913$ across $L \in \{1,2,3,5,10\}$, while Recall@FPR1\% remains in the range from $0.845$ to $0.866$. Increasing circuit depth therefore does not produce a clear improvement for angle encoding. QLR-amplitude improves from $L=1$ to $L=3$ and then stabilises. At $N=1000$, PR-AUC increases from $0.695$ at $L=1$ to $0.775$ at $L=3$, but the gap with QLR-angle remains large. QLR-DR reaches its best discrimination performance at $L=1$ and decreases as $L$ increases. At $N=1000$, PR-AUC decreases from $0.912$ at $L=1$ to $0.887$ at $L=10$, while Recall@FPR1\% decreases from $0.852$ to $0.828$. For QLR-DR, increasing depth also reintroduces the input features more times and adds more trainable operations. Therefore, in the multi-qubit data-reuploading implementation on \texttt{HTRU-$2$}, added depth does not improve PR-AUC or Recall@FPR1\%. However, this result should not be considered a general limitation of data re-uploading, but a property of the present architecture, dataset, and training setup.

The depth-scaling results show that deeper circuits are not automatically beneficial in this setting. QLR-angle is already effective at small depth. QLR-amplitude benefits from a small increase in depth but remains weaker overall. For QLR-DR, larger depth is associated with lower discrimination performance over the tested range. Across the tested configurations, encoding choice has a larger effect on discrimination performance than increasing circuit depth.

\begin{figure*}[!t]
\centering
\includegraphics[width=\linewidth]{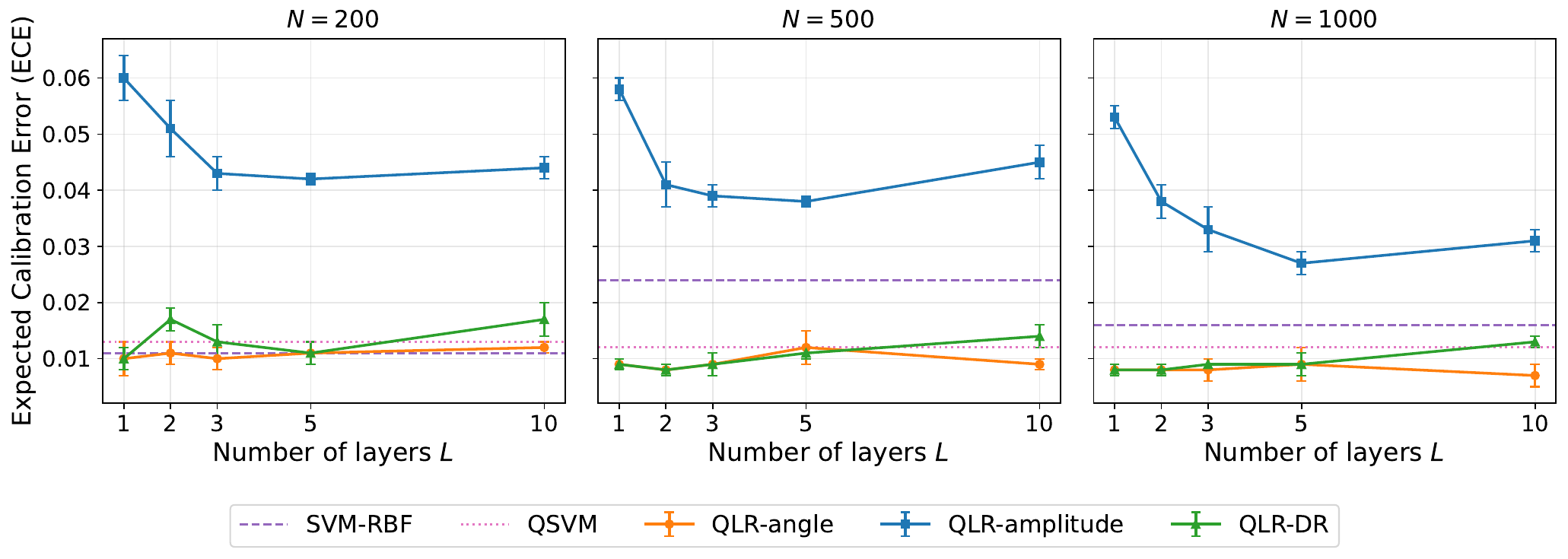}
\caption{ECE as a function of circuit depth for all QLR variants and $N \in \{200,500,1000\}$. Error bars denote standard error over three independent seeds. Horizontal reference lines indicate the SVM-RBF and QSVM values for the corresponding training size.
}
\label{fig:ece_depth}
\end{figure*}

\Cref{fig:ece_depth} reports ECE as a function of circuit depth. This figure provides a first view of how calibration error changes across the same depth and training-size settings. The horizontal reference lines indicate the SVM-RBF and QSVM ECE values for the corresponding training size. SVM-RBF is shown as the strongest classical calibration baseline at the benchmark configuration.

QLR-angle maintains low and stable ECE across the tested depths and training sizes. QLR-DR also has low ECE at small depth, but its calibration error increases as the depth becomes larger. QLR-amplitude starts with higher ECE and improves slightly with depth, but it remains less stable than QLR-angle. For QLR-DR, the increase in ECE occurs in the same depth range where PR-AUC and Recall@FPR1\% decrease. A more detailed calibration analysis using reliability diagrams and Murphy decomposition is given in \cref{subsec:calibration}.

\begin{figure*}[!t]
\centering
\begin{subfigure}[t]{0.48\textwidth}
    \centering
    \includegraphics[width=\linewidth]{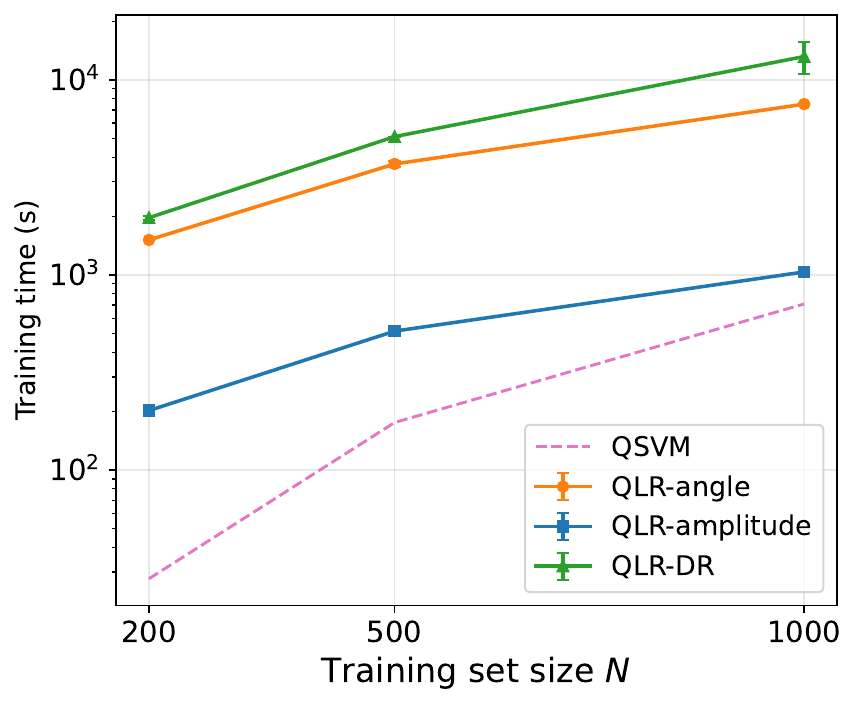}
    \caption{Training time versus dataset size at $L=3$}
    \label{fig:runtime_data}
\end{subfigure}
\hfill
\begin{subfigure}[t]{0.48\textwidth}
    \centering
    \includegraphics[width=\linewidth]{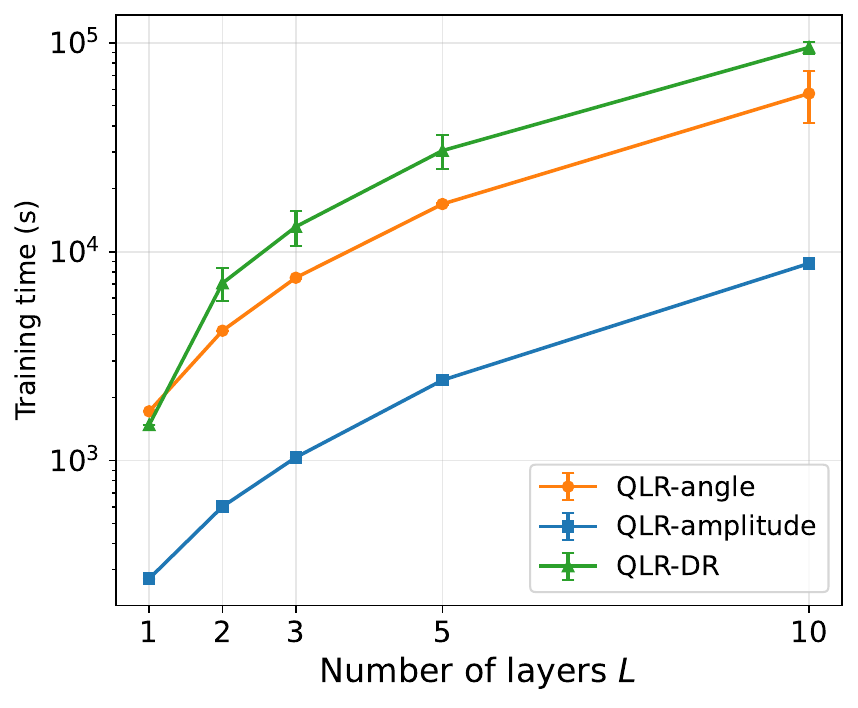}
    \caption{Training time versus circuit depth at $N=1000$}
    \label{fig:runtime_depth}
\end{subfigure}
\caption{
Training-time scaling for the QLR variants. The left panel shows training time as a function of training-set size at fixed depth $L=3$. The right panel shows training time as a function of circuit depth at fixed training size $N=1000$. 
QSVM training time is included as a reference in the left panel. The vertical axes use logarithmic scaling.
}
\label{fig:runtime}
\end{figure*}

\Cref{fig:runtime} reports the training-time cost associated with the scaling experiments. The left panel shows training time as a function of training-set size at fixed depth $L=3$. The right panel shows training time as a function of circuit depth at fixed training size $N=1000$. QSVM training time is included in the left panel as a reference.

Training time increases with both $N$ and $L$. QLR-amplitude has the lowest training cost among the QLR variants across all configurations. At $N=1000$ and $L=3$, QLR-amplitude requires approximately $1{,}036$ seconds, compared with nearly $7{,}505$ seconds for QLR-angle and $13{,}170$ seconds for QLR-DR. The higher cost of QLR-DR follows from the repeated data encoding at each layer, so the number of data-dependent operations increases with circuit depth. At $N=1000$ and $L=10$, QLR-DR requires approximately $95{,}000$ seconds, compared with roughly $57{,}000$ seconds for QLR-angle and $8{,}800$ seconds for QLR-amplitude.

QSVM training time is lower than QLR-angle and QLR-DR for the dataset sizes considered, but its prediction cost is much larger. As reported in \cref{tab:calibration_runtime}, QSVM prediction time exceeds $5{,}000$ seconds at $N=1000$, while all QLR variants remain below $12$ seconds. The QLR models are therefore expensive to train in simulation but comparatively fast at inference. Increasing depth or training-set size gives only limited performance gains relative to the additional cost. For the \texttt{HTRU-$2$} dataset, shallow QLR-angle circuits provide the clearest balance between discrimination, calibration, and runtime among the QLR variants tested.

\subsection{Probability Calibration}
\label{subsec:calibration}

This section examines the calibration of the probability estimates produced by the QLR variants and relates these estimates to pulsar candidate ranking. Discrimination metrics measure ranking performance, but pulsar candidate selection also requires probability estimates that are meaningful for follow-up decisions. Calibration is analysed using the Brier score, ECE, reliability diagrams, and Murphy decomposition, as defined in \cref{subsec:metrics}. The analysis compares numerical calibration metrics with visual reliability diagnostics and separates probability reliability from class separation through Murphy decomposition.

Calibration across circuit depth is first examined using reliability diagrams.

\begin{figure*}[t]
\centering
    \includegraphics[width=\linewidth]{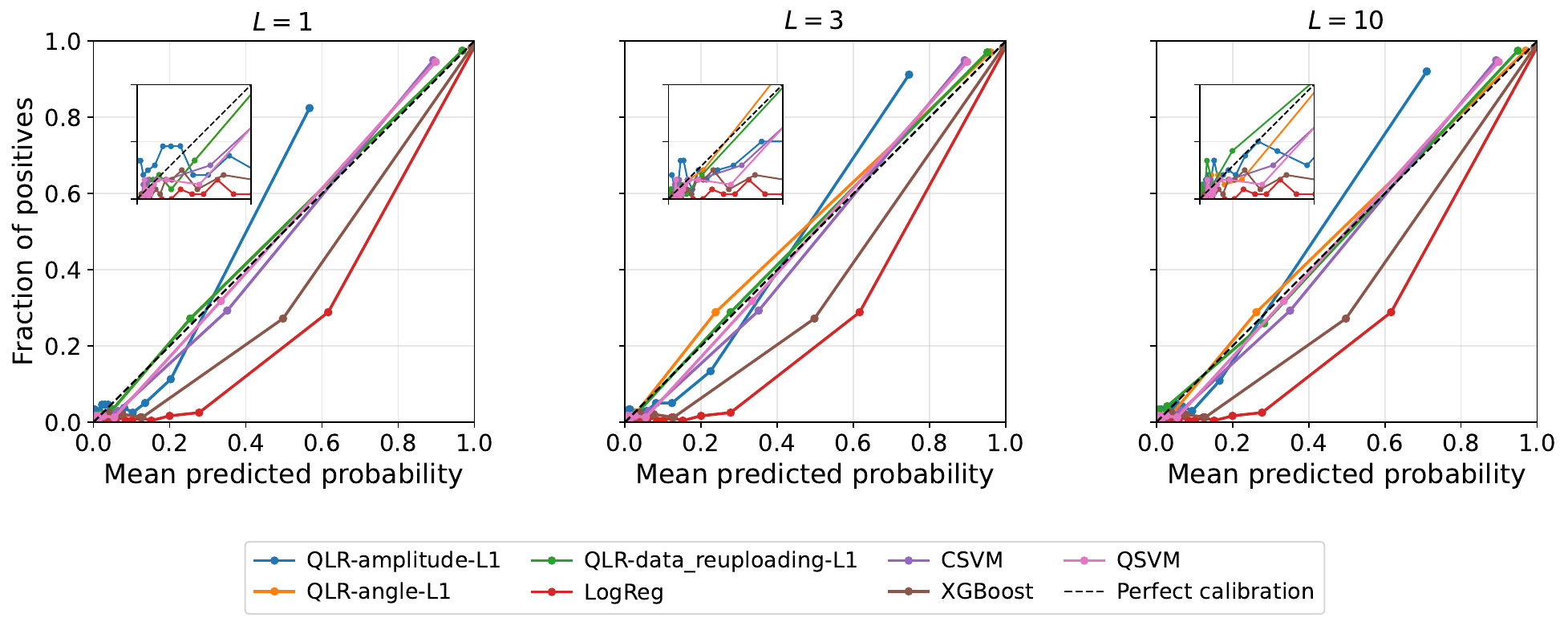}
\caption{
Reliability diagrams at $N=1000$ for circuit depths $L \in \{1,3,10\}$. 
Each curve compares the mean predicted pulsar probability with the observed fraction of positives in probability bins. The dashed diagonal represents perfect calibration. Insets show the low-probability region, where many non-pulsar candidates are expected to lie. Curves are shown only for occupied probability bins. For QLR-angle at $L=1$, several occupied bins lie close to the low-probability region and are visually compressed even in the inset.
}
\label{fig:reliability_depth}
\end{figure*}

\Cref{fig:reliability_depth} shows reliability diagrams at fixed dataset size $N=1000$ for three representative depths, $L \in \{1,3,10\}$. Each panel compares the mean predicted probability with the observed fraction of positives in probability bins. Curves close to the diagonal indicate better calibration, while deviations from the diagonal indicate a mismatch between predicted probabilities and empirical frequencies. The inset shows the low-probability region, where many non-pulsar candidates are expected to lie. For QLR-angle at $L=1$, several occupied bins lie very close to this low-probability region, so the curve is visually compressed even in the inset. 
The reliability diagrams are therefore interpreted together with the Brier score, ECE, and Murphy reliability term.

The behaviour of the curves in \cref{fig:reliability_depth} is consistent with the numerical calibration values in \cref{tab:calibration_runtime}. At $N=1000$ and $L=3$, QLR-angle has the lowest Brier score and ECE among the QLR variants, with Brier $0.018 \pm 0.001$ and ECE $0.008 \pm 0.002$. Its reliability curve remains close to the diagonal, supporting the low numerical calibration error reported in \cref{tab:calibration_runtime}. QLR-DR is close to QLR-angle in the table, with Brier $0.019 \pm 0.001$ and ECE $0.009 \pm 0.000$, and its reliability curves also remain close to the diagonal for the displayed depths. For QLR-DR, ECE increases with depth even though the reliability curves remain visually close to the diagonal. This can occur when several probability bins show small deviations between predicted probabilities and observed frequencies, producing a higher average calibration error without a large visual deviation in the reliability diagram. QLR-amplitude has higher ECE at $L=3$ and shows larger fluctuations across probability bins, consistent with its less stable reliability curves.

The depth-dependent calibration trends show that increasing circuit depth does not consistently improve probability estimates. QLR-angle remains stable as depth increases. QLR-DR remains visually close to the reliability diagonal, but its ECE increases at larger $L$. QLR-amplitude improves slightly with depth but remains less stable than QLR-angle. For QLR-DR, the increase in ECE occurs in the same depth range where PR-AUC and Recall@$\mathrm{FPR}=1\%$ decrease, as shown in \cref{fig:depth_scaling}.

Murphy decomposition is then used to examine whether the calibration and Brier-score trends are mainly associated with probability reliability or class separation.

\begin{figure*}[t]
\centering
\includegraphics[width=\textwidth]{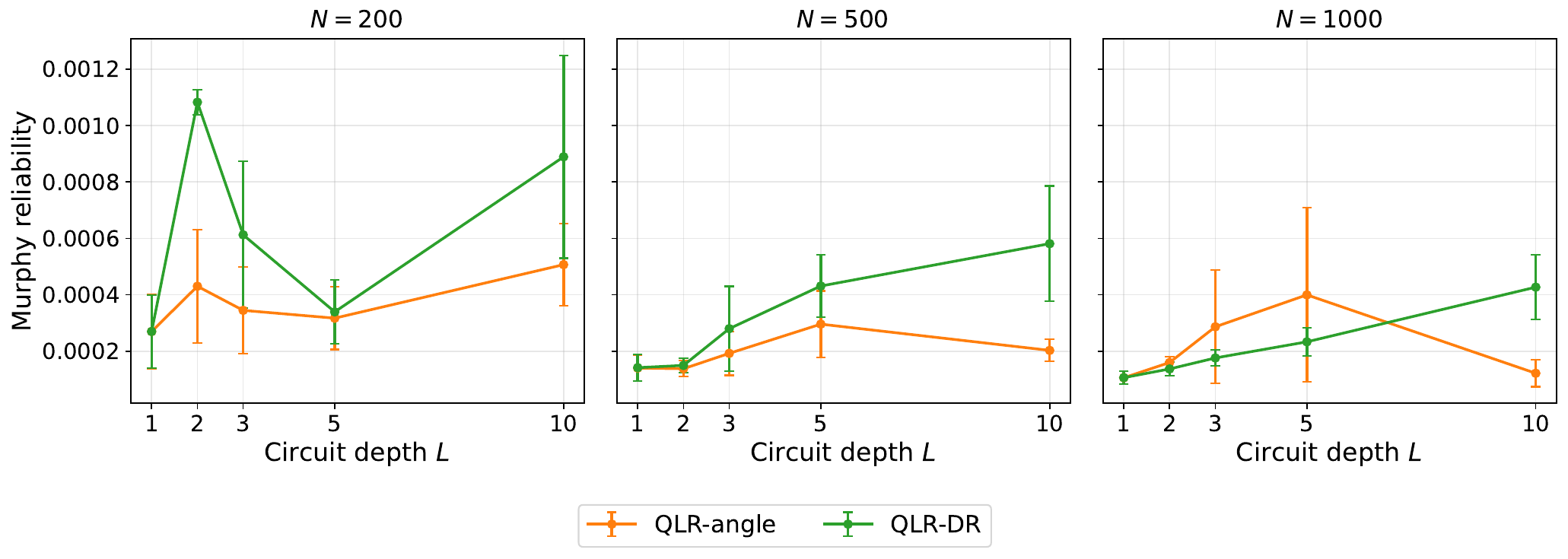}
\caption{
Murphy reliability term as a function of circuit depth for QLR-angle and QLR-DR. 
Each panel corresponds to a training-set size $N \in \{200,500,1000\}$. 
Lower values indicate better agreement between predicted probabilities and observed frequencies. 
The minimum possible value is zero. Error bars denote standard error over three independent seeds.
}
\label{fig:murphy_reliability}
\end{figure*}

\begin{figure*}[t]
\centering
\includegraphics[width=\textwidth]{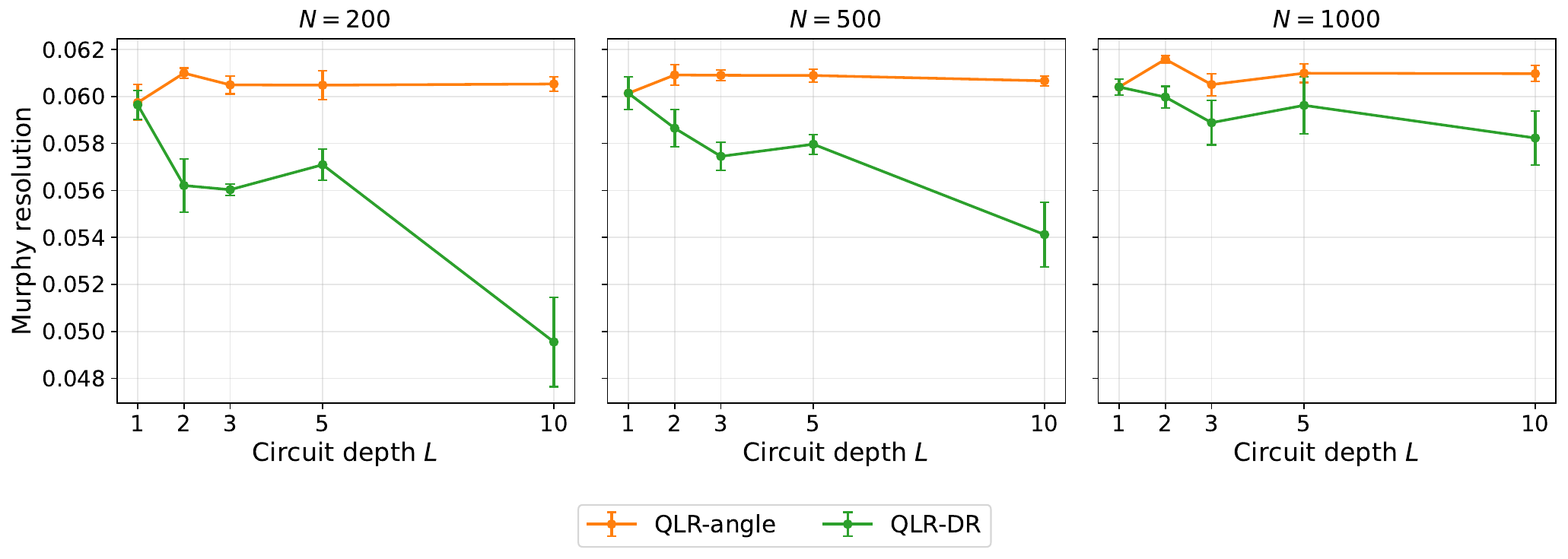}
\caption{
Murphy resolution term as a function of circuit depth for QLR-angle and QLR-DR. 
Each panel corresponds to a training-set size $N \in \{200,500,1000\}$. Higher values indicate better separation between groups of candidates that contain different proportions of true pulsars. 
The resolution term is interpreted relative to the uncertainty term $\bar{y}(1-\bar{y})$, where $\bar{y}$ is the positive-class prevalence. Error bars denote standard error over three independent seeds.
}
\label{fig:murphy_resolution}
\end{figure*}

\Cref{fig:murphy_reliability} reports the Murphy reliability term, while \cref{fig:murphy_resolution} reports the corresponding resolution term for QLR-angle and QLR-DR across circuit depths and training-set sizes. 
The reliability term is minimised at zero, where predicted probabilities match observed frequencies within the calibration bins. The resolution term is bounded above by the uncertainty term $\bar{y}(1-\bar{y})$, where $\bar{y}$ is the positive-class prevalence. 
For the \texttt{HTRU-$2$} dataset used here, $\bar{y}\approx 0.09$, giving an uncertainty value of approximately $0.082$.

The reliability term remains small across depths, with values around $10^{-4}$ to $10^{-3}$. 
These values are small relative to the Brier-score scale set by the uncertainty term. 
However, some points have standard errors that are large relative to the mean, especially for QLR-angle at $N=1000$, $L=5$, and for QLR-DR at smaller training sizes. 
These points should therefore not be interpreted as evidence of a stable depth-dependent reliability trend. 
The larger standard errors indicate that the bin-level calibration estimate varies across seeds. 
This variability is plausible under class imbalance, because some probability bins may contain few positive examples. In addition, only three independent seeds are used in this study, which limits the precision of the standard-error estimates. A larger number of repeated runs would provide more stable uncertainty estimates for the Murphy reliability term.

The resolution term shows a clearer depth-dependent pattern. QLR-angle maintains resolution values around $0.06$ across depths, corresponding to a substantial fraction of the uncertainty value. 
This shows that QLR-angle preserves a substantial fraction of the available resolution. 
QLR-DR shows a stronger decrease in resolution as depth increases. 
Thus, the depth-dependent degradation observed for QLR-DR is more visible in the resolution term than in the reliability term. In this setting, deeper reuploading circuits mainly reduce class separation, rather than producing a large and stable increase in calibration error.
One possible interpretation is the structure of the learned score before the logistic mapping. 
In the QLR framework, the logistic layer operates on features $z(x;\theta)$ obtained from Pauli-$Z$ expectation values, with each component bounded in the interval $[-1,1]$. 
Together with regularisation of the logistic weights, this bounded feature range can limit extreme values of the linear score $w^\top z(x;\theta)+b$ before the logistic function is applied. 
This may reduce overconfident probability estimates. QLR-angle combines this bounded feature representation with a shallow circuit, giving stable score distributions across the tested configurations. QLR-DR introduces repeated input encoding and additional trainable layers. 
In this implementation, larger depth is associated with more variable optimisation and lower resolution at larger $L$. The comparison should therefore be considered as an implementation-specific result, rather than as evidence that multi-qubit data re-uploading is generally less well calibrated.

\section{Conclusion}\label{sec:conclusion}

This work studied quantum logistic regression for pulsar candidate classification under class imbalance. Three encoding strategies were compared: angle encoding, amplitude encoding, and data re-uploading. The analysis considered discrimination performance, probability calibration, and computational cost across different circuit depths and training-set sizes. 
Classical baselines and a QSVM model were included for comparison.

For the \texttt{HTRU-$2$} setting studied here, shallow QLR circuits achieve competitive discrimination performance. QLR-angle gives the best overall performance among the QLR variants, with limited variation across circuit depth and training-set size. 
Its performance remains close to the best classical baselines in the low-false-positive-rate regime. Increasing circuit depth beyond a small number of layers does not improve performance for QLR-angle. QLR-DR performs well at small circuit depth, but its discrimination performance decreases as depth increases in the present multi-qubit implementation. QLR-amplitude improves with depth but remains below the other QLR variants across the tested configurations.

Calibration analysis shows clear differences between encodings. QLR-angle produces stable and well-aligned probability estimates across circuit depths. QLR-DR is well calibrated at small depth, but its ECE increases at larger depth. QLR-amplitude shows larger fluctuations and remains less stable. Murphy decomposition gives additional context for these trends. For QLR-angle, the reliability term remains close to zero, while the resolution term remains high relative to the uncertainty scale of the dataset. The low Brier score is therefore mainly associated with strong class separation, with only a small contribution from bin-level calibration error. 
For QLR-DR, the main depth-dependent change appears in the resolution term rather than in a large systematic increase in reliability error.

The calibration differences between models may be related to how prediction scores are mapped to probabilities. In QLR, the logistic layer acts on bounded expectation-value features, which can limit extreme scores before the logistic mapping when combined with regularisation. 
QLR-angle combines this bounded feature representation with a shallow circuit, giving stable score distributions in the tested configurations. Deeper QLR-DR circuits introduce repeated input encoding and additional trainable layers. In this implementation, larger depth is associated with more variable optimisation and less stable score distributions.

From a practical perspective, the results show a trade-off between discrimination, calibration, and computational cost. QLR-angle gives the clearest balance among the QLR variants tested, with strong discrimination and stable calibration at lower cost than QLR-DR. QLR-DR has higher computational cost without consistent performance gains at larger depth. QSVM achieves competitive discrimination, but its prediction cost becomes large as the dataset size increases. 
The favourable calibration behaviour observed for some quantum models should therefore be interpreted alongside their runtime cost, which remains a practical limitation for large-scale deployment.

For pulsar discovery, well-calibrated probabilities are critical because candidate ranking and telescope follow-up decisions depend on reliable confidence estimates. 
A classifier that ranks candidates well but assigns poorly calibrated probabilities may still be difficult to use in follow-up pipelines. 
Reliable probability estimates are therefore central to the efficient allocation of observational resources.

This study is limited to noiseless simulation, one dataset, and three independent seeds. 
The paired-seed protocol controls for variation due to data splits and initialisation, but the reported standard errors should be interpreted as descriptive uncertainty estimates rather than as the basis for formal statistical inference. 
Future work should increase the number of seeds and apply paired statistical tests, such as the Wilcoxon signed-rank test, to assess the robustness of the observed differences. 
Further work should also evaluate these models under realistic hardware constraints, including noise, entangling-gate overhead, and trainability limitations on near-term quantum devices \cite{preskill2018quantum,larocca2025barren}. Additional datasets should be considered to assess whether the same calibration and performance trends persist beyond pulsar candidate classification. Further investigation of encoding strategies and training methods may help improve stability and scalability in deeper circuits.

\section*{Acknowledgements}

The authors thank Charles Varmantchaonala Moudina for helpful discussions in early stages of the work.  CMM  acknowledges the financial support made possible by a grant from Carnegie Corporation of New York (provided through the African Institute for Mathematical Sciences Research and Innovation Centre and Quantum Leap Africa). The statements made and views expressed are solely the responsibility of the authors.

\section{Author Contributions}

CMM conceived and designed the study. CMM performed the numerical experiments and analysed the results. CMM prepared the manuscript. DS contributed to methodological refinement and manuscript review. PKO and FP supervised the research. All authors read and approved the final manuscript.

\section{Conflicts of interests}

All authors declare no conflicts of interests.

\bibliography{bibfile}

\end{document}